\documentclass[10pt,letterpaper,twocolumn]{article}

\usepackage{ol2}
\usepackage[draft]{hyperref}

\usepackage{amsmath}
\usepackage{graphicx}

\usepackage{hyperref}

\newcommand{\be}{\begin{equation}}
\newcommand{\ee}{\end{equation}}

\newcommand{\beq}{\begin{equation}}
\newcommand{\eeq}{\end{equation}}
\newcommand{\beqa}{\begin{eqnarray}}
\newcommand{\eeqa}{\end{eqnarray}}

\newcommand{\ben}{\begin{enumerate}}
\newcommand{\een}{\end{enumerate}}
\newcommand{\bit}{\begin{itemize}}
\newcommand{\eit}{\end{itemize}}
\newcommand{\bpm}{\begin{pmatrix}}
\newcommand{\epm}{\end{pmatrix}}

\newcommand{\PT}{\mathcal{PT}}
\newcommand{\p}{\mathcal{P}}
\newcommand{\T}{\mathcal{T}}
\newcommand{\tE}{\tilde{E}}
\newcommand{\tb}{\tilde{\beta}}

\newcommand{\eref}[1]{Eq.~(\ref{#1})}

\begin{document}

\twocolumn[

\title{Discrete vortex solitons and $\PT$ symmetry}

\author{Daniel Leykam$^1$, Vladimir V. Konotop$^2$, and Anton S. Desyatnikov$^1$}

\address{
$^1$Nonlinear Physics Centre, Research School of Physics and Engineering\\
The Australian National University, Canberra ACT 0200, Australia\\
$^2$Centro de F\'{\i}sica Te\'orica e Computacional and Departamento de F\'{\i}sica\\
Faculdade de Ci\^encias, Universidade de Lisboa, Lisboa 1649-003, Portugal
}

\begin{abstract}
We study the effect of lifting the degeneracy of vortex modes with a $\PT$ symmetric defect, using discrete vortices in a circular array of nonlinear waveguides as an example. When the defect is introduced, the degenerate linear vortex modes  spontaneously break $\PT$ symmetry and acquire complex eigenvalues, but nonlinear propagating modes with real propagation constants can still exist. The stability of nonlinear modes depends on both the magnitude and the sign of the vortex charge, thus $\PT$ symmetric systems offer new mechanisms to control discrete vortices.
\end{abstract}

]

Nonlinear periodic media allow for efficient localization and routing of light signals in the form of discrete optical solitons~\cite{disc_opt}. Alongside usual signals encoded in a soliton's amplitude, the phase of a vortex soliton also carries a quantized bit of information, its topological charge $m$ (TC)~\cite{review}. The absolute value of TC, $|m|$, is the phase winding number, in units of $2\pi$, around the vortex origin, and the sign of TC determines the direction of power flow, or vortex handedness. Periodic media support stable discrete vortex solitons~\cite{malomed2001} and allow for robust control over vortex TC by switching between integer values $m\leftrightarrow -m$, known as charge flipping~\cite{alexander2004,bezryadina2006,Desyatnikov_PRA}.

In systems with discrete rotational symmetry, such as a ring of $N$ identical waveguides~\cite{Desyatnikov_PRA}, vortex and anti-vortex modes with phase $\sim\exp(i2\pi m n/N)$ are degenerate, here $n=1,2\dots N$ is the waveguide number. The charge-flipping transformation $m\leftrightarrow -m$ can be seen as a complex conjugation, or time-reversal operation. It is interesting to explore the additional possibilities for TC control which can be offered by dissipative parity-time ($\PT$) symmetric systems~\cite{Bender_PRL}, which have pure real spectra below the $\PT$-symmetry breaking point. Recent studies of infinite chains and discrete rings of coupled optical waveguides with gain and loss have demonstrated their direct relevance to applications~\cite{BinaryLattice,nature,Li_PRE,ZezKon_PRL,Sukhorukov_OL}. The existence of propagating modes with real spectra in such arrays readily suggests the possibility of the existence of propagating vortices. Broken $\T$ symmetry implies that vortices with opposite charges can behave differently, i.e. their degeneracy is lifted.

In this Letter we show that nonlinear propagating vortex modes can still exist even when the $\PT$ symmetry breaking threshold is reached for the linear system. We explore interesting consequences of lifted vortex degeneracy. The onset of modulational instability becomes sensitive to both the magnitude and sign of vortex charge, thus expanding the available toolbox for TC control. The sensitivity to the sign of vortex charge is impossible in systems respecting $\T$ symmetry.

\begin{figure}
\includegraphics[width=\columnwidth]{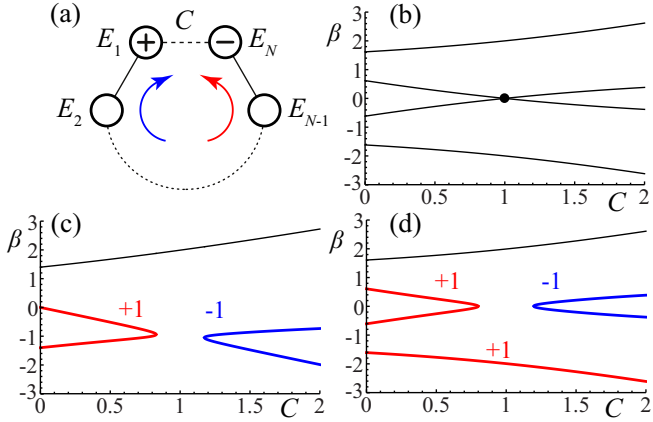}
\caption{(Color online) (a) Schematic of $N$-site ring with gain (+) and loss (-) at waveguides $1$ and $N$. Phase circulation direction of vortices with $m>0$ and $m<0$ is indicated by anti-clockwise (red) and clockwise (blue) arrows. (b) 
$\beta$ {\it vs.} $C$ for a conservative ($\gamma = 0$) $N=4$ ring. Degenerate vortex modes 
only  occur at the intersection marked by the black circle. (c) and (d) Linear spectrum for $N=3$ and $N=4$ rings with $\gamma = 0.2$. Modes with TCs +1 (-1) are shown in red (blue). 
}
\label{fig:linearmodes}
\end{figure}

We consider a circular array of $N$ waveguides with Kerr nonlinearity, shown in
Fig.~\ref{fig:linearmodes}(a), with gain and loss located at waveguides 1 and $N$, coupling $C$ between them, and nearest neighbor coupling normalized to unity along the rest of the ring (the difference in the coupling constants can be introduced by variation of the distances between the respective waveguides),
\begin{eqnarray}
&&i \partial_z E_1 + C E_N + E_2 - i \gamma E_1 + \delta | E_1|^2 E_1 = 0, \nonumber\\
&&i \partial_z E_{n} + E_{n-1} + E_{n+1} + \delta |E_n|^2 E_n = 0,\label{main}\\
&&i \partial_z E_N + C E_1 + E_{N-1} + i \gamma E_N + \delta | E_N |^2 E_N = 0.\nonumber
\end{eqnarray}
Here $E_n(z)$ ($n=1...N$) is a dimensionless electric field in the $n-$th waveguide, $E_{n+N} = E_n$, and $\delta = \pm 1$ is the nonlinear coefficient. Stationary modes take the form $E_n(z)\to E_n \exp(i \beta z)$, where $\beta$ is the propagation constant. Aspects of the linear, large $N$ limit of~\eref{main} were previously considered in Refs.~\cite{Scott_PRA,Sukhorukov_OL}.
In discrete conservative systems the TC is $m = \frac{1}{2 \pi}\sum_{n=1}^N \text{Arg}[ E_n^* E_{n+1} ]$, and measures the phase winding along
the contour. $J_{n} = 2 \mathrm{Im}(E_n^* E_{n+1})$ is the energy flow from site $n$ to site $n+1$~\cite{Desyatnikov_PRA}. The usual assumption is that nonzero TC ($m > 0$) $m < 0$ corresponds to the (anti)clockwise circulation of phase and energy around the contour. This definition breaks down for even $N$ at the band edge $|m|=N/2$, which are multipole modes $E_n\sim (-1)^n$ without energy circulation (ie. not a vortex).

The equation for stationary solutions in ring-type systems can be cast in the general form
\be
\label{general}
-\beta E  + HE+\delta F(|E|^2)E  = 0, \qquad H=H_0 + i\gamma H_{1}
\ee
where $E=(E_1,...,E_N)^T$ is a column vector ($T$ stands for transposition) and the nonlinearity is given by the diagonal matrix $F(|E|^2)=$diag$(|E_1|^2,...,|E_N|^2)$. Linear operator $H_0$ is a matrix describing the array without dissipation and $H_1$ describes the losses. One ensures that the commutator $[H_0,\mathcal{P}] = 0$ and $H_1\mathcal{P} = -\mathcal{P} H_1$ with the ``parity'' inversion matrix $\p$ having only anti-diagonal nonzero elements, $\p_{ij}=\delta_{i,N+1-j}$, and that $[H_{0,1},\mathcal{T}] = 0$, where $\T$ is the operator of complex conjugation. Thus $[H,\mathcal{PT}] = 0$, i.e. $H$ is $\PT$-symmetric.

First, we show that if $H_0$ supports at least one degenerate pair of vortex modes $E_\pm$ belonging to a double degenerate eigenvalue $\beta_0$, i.e. $H_0E_\pm=\beta_0E_\pm$, and $H_1$ is symmetric, such as in~\eref{main}, the $\PT$ symmetry breaking threshold $\gamma_{th}$, i.e. the value of $\gamma$ above which the spectrum has complex eigenvalues, is zero: $\gamma_{th}=0$. Indeed, from the above properties of $H_0$ it follows that one can choose $E_\pm=\p E_\mp=\T E_\mp=E_\mp^*$. Hence $\tE^{(1)}=E_++E_-$ and $\tE^{(2)}=i(E_+-E_-)$ are the two real eigenstates of $H_0$ and $\p \tE^{(j)}=(-1)^{j+1}\tE^{(j)}$ ($j=1,2$). Then, from the Theorem~2.1 of Ref.~\cite{Caliceti_JPhysA2006} follows that $H$ has a pair of complex eigenvalues for arbitrarily small $\gamma$.

The symmetry breaking threshold 
of~\eref{main} at $C\neq 1$ is non-zero~\cite{Sukhorukov_OL}. Hence, there are no vortex eigenstates of $H_0$. Instead, vortex eigenstates appear at nonzero $\gamma <\gamma_{th}$. 
Moreover, the $\PT$ operator does not change TC, i.e. if $\tE$ is an eigenmode of $H$ with the charge $m$, $H\tE=\tb\tE$ at $\gamma<\gamma_{th}$, then $\PT \tE$ is also an eigenmode with the same $\tb$ and $m$, and thus one can set $\PT \tE=\tE$. In other words one can consider linear $\PT$-symmetric solutions.
In Fig.~\ref{fig:linearmodes}(b) we show the spectrum for an $N=4$ ring when $\gamma = 0$. Vortex modes only exist at $C = 1$, and they are degenerate.
The effect of nonzero $\gamma$ is shown in Figs.~\ref{fig:linearmodes}(c,d) for odd $N=3$ with  $\gamma_{th}=(C^2-3C^{2/3}+2)^{1/3}$ and even $N=4$ with $\gamma_{th}=|C-1|$. The degeneracy becomes a pair of exceptional points, and branches with nonzero topological charge $|m| = 1$ appear.
The breaking of $\T$ symmetry manifests itself through the separation of different charges into distinct branches. Curiously, nonzero $m$ here does not imply nonzero vorticity: the multipole (lowest $\beta$) mode in Fig.~\ref{fig:linearmodes}(d) has $m=1$, but it is not a vortex: there is phase winding, but its phase does not increase monotonically, resulting in flow from the gain site to the dissipative site along both paths. In contrast the $|m|=1$ modes which vanish as $C \rightarrow 1$ are true vortices with energy circulation (the sign of $J_n$ is the same for all $n$).

\begin{figure}
\includegraphics[width=\columnwidth]{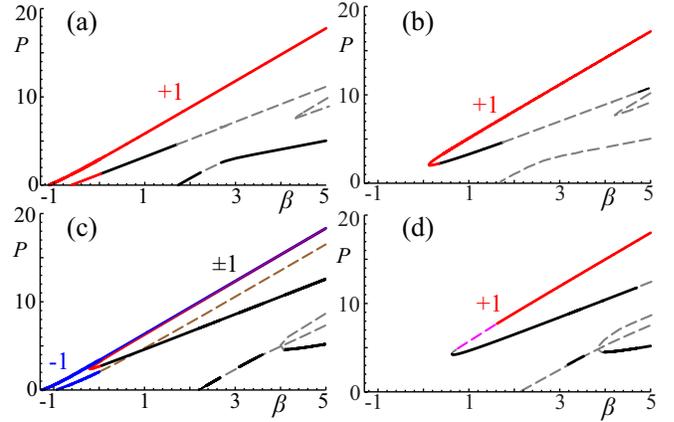}
\caption{(Color online) Stable (unstable) nonlinear modes of $N=3$ ring are shown with solid (dashed) lines. 
TCs are indicated next to the curves, $m= +1$ in red (purple), $m=-1$ in blue (brown), and $m=0$ in black (grey). Parameters are: (a) $C = 0.6, \gamma = 0.2<\gamma_{th}$; (b) $C= 0.6, \gamma = 0.6>\gamma_{th}$; (c) $C = 1.3, \gamma = 0.2<\gamma_{th}$; and (d) $C = 1.3, \gamma = 0.6>\gamma_{th}$.}
\label{fig:nonlinearmodes}
\end{figure}

Since the nonlinearity in~\eref{general} has the symmetry $\PT \left(F(|\tE|^2)\tE\right)=F(|\tE|^2)\tE$ there exist~\cite{ZezKon_PRL} nonlinear propagating vortex modes bifurcating from each of the linear $\PT$-symmetric vortices. These modes do not exhaust all possible soliton solutions with real propagation constants which can exist even when $\PT$ symmetry is broken in the linear regime~\cite{ZezKon_PRL}. In Fig.~\ref{fig:nonlinearmodes} we present families $P(\beta)$  of nonlinear modes obtained using Newton's method (here $P=\sum_{n=1}^N|E_n|^2$ is the total power). Note that for $\gamma\neq 0$ the total power is not conserved except for stationary modes. We consider below the simplest $N=3$ ring in detail, then discuss properties of larger $N$ systems.

Fig.~\ref{fig:nonlinearmodes}(a) shows the typical $N=3$ soliton family for $C < 1$ with $\gamma < \gamma_{th}$. We observe the bifurcation of $m=1$ vortices from linear modes at $P=0$ discussed in Fig.~\ref{fig:linearmodes}(c). As the power is increased, the energy flow between $E_1$ and $E_N$ in the lower branch decreases and eventually changes sign, destroying the vortex, and the charge becomes zero. No $m=-1$ modes are found in the parameter range scanned: there are no such linear modes, and the broken $\T$ symmetry suppresses their saddle-node bifurcation.

When $\gamma$ is increased beyond $\gamma_{th}$ in Fig.~\ref{fig:nonlinearmodes}(b), the linear modes merge and annihilate, and pairs of $m=+1$ vortices now appear at a saddle-node bifurcation (notice that the unstable branch is not visible on the scale of the figure). One branch quickly loses its charge and vorticity; this happens at lower powers as $\gamma$ is increased. Once again, no $m=-1$ modes are found.

The situation for $C>1$,
[Figs.~\ref{fig:nonlinearmodes}(c,d)] is different. Now nonlinear $m=-1$ modes exist below $\gamma_{th}$, because they can bifurcate from linear modes. In this case one is stable at high power, while the other is unstable, but both maintain their vorticity. A pair of $m=+1$ vortices is still created at a saddle node bifurcation,
with behavior similar to Figs.~\ref{fig:nonlinearmodes}(a,b);
 one branch loses its charge and vorticity, while the other remains stable. Above the $\mathcal{PT}$ breaking threshold, the $m=-1$ modes are destroyed, while the $m=+1$ saddle node bifurcation remains.

Evidently, the $\mathcal{PT}$ symmetric defect in combination with the asymmetric coupling $C$ determines the charge of linear vortex modes - the energy flows from the site with gain to the site with loss along the path with stronger coupling, and back via the path with weaker coupling. In contrast, the charge of nonlinear modes above $\gamma_{th}$ is mainly sensitive to the distribution of gain and loss. These charges are opposite when $C > 1$.

As $N$ is increased modes with larger $|m|\le N/2$ appear. In conservative systems with focusing nonlinearity the modes with higher $|m|$ are stable, while lower charged vortices suffer instabilities above a critical power~\cite{Desyatnikov_PRA}. This behavior holds in $\PT$ symmetric rings, but the sign of the charge also plays an important role. For even $N$ there is an additional nonlinear branch bifurcating from the linear multipole mode, which formally has the highest TC $|m|=N/2-1$ supported by the ring, but it does not form a vortex. 

\begin{figure}
\includegraphics[width=\columnwidth]{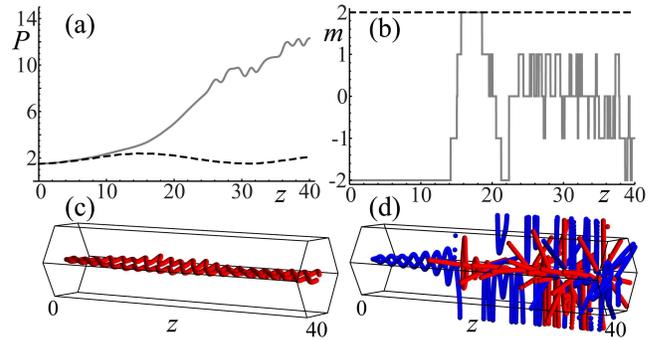}
\caption{(Color online) Charge selectivity in an array with $N=6$, $C=1$, $\gamma =0.2$, and $\delta = 1$. The total power (a) and TC (b) are shown vs. propagation distance $z$ for $m=+2$ (black, dashed) and $m=-2$ (gray, solid) inputs. Vortex (red) and antivortex (blue) lines are shown in (c) for input $m=+2$ and in (d) for $m=-2$.}
\label{fig:selectivity}
\end{figure}

As an example of nonlinear vortex dynamics, we consider a hexagonal ($N=6$) coupler, with $C=1$ and $\delta=1$. When $\gamma = 0$, the symmetric vortex modes of charge $m$ take the form $E_n = A \exp( 2 \pi i m n/N + i \beta z)$~\cite{Desyatnikov_PRA} and their stability is independent of the vorticity direction, ${\rm sign}(m)$. The situation is different with $\T$ symmetry broken, as we demonstrate in Fig.~\ref{fig:selectivity} for $\gamma = 0.2$ and $A=0.5$, by contrasting the propagation of $m=\pm 2$ vortices. The $m=+2$ perturbed vortex oscillates stably about a nonlinear vortex mode, and its charge is conserved. In contrast, the $m=-2$ input experiences $\PT$ breaking and its power grows exponentially. These dynamics are further illustrated by obtaining vortex lines from the array supermode (cf. Ref.~\cite{Desyatnikov_PRA}), shown in Figs.~\ref{fig:selectivity}(c,d). We see the stable oscillation of a pair of charge 1 vortices for the $m=+2$ input, while the opposite input charge $m=-2$ results in highly irregular vortex dynamics. Inputs with $|m|=1$ are also unstable. With repulsive nonlinearity $\delta = -1$ the stability inverted: the $m=1$ vortex is stable, while all others are unstable. In this example the nonlinearity is essential: $\PT$ breaking occurs in the linear limit, and vortex inputs of any charge are decomposed into amplified and attenuated components.

In conclusion, we have studied discrete vortices in $N$-site rings of coupled nonlinear waveguides with a $\PT$ symmetric defect. We have shown that by breaking $\T$ symmetry, the existence, stability and dynamics of nonlinear vortex modes become sensitive to the sign of their charge, offering an additional degree of freedom for all-optical control of discrete vortices.

This work was supported by the Australian Research Council. The work of
 VVK was supported by the FCT (Portugal) grants PTDC/FIS/112624/2009 and PEst-OE/FIS/UI0618/2011.

\pagebreak

\end{document}